\documentclass[epj,twocolumn]{webofc}
\usepackage[varg]{txfonts}   
\newcommand{\be}{\begin{equation}}
\newcommand{\ee}{\end{equation}}
\newcommand{\mh}{M_{\rm h}}
\newcommand{\rt}{R_{\rm t}}
\newcommand{\rp}{R_{\rm p}}
\newcommand{\rs}{R_{\rm S}}
\newcommand{\mdot}{\dot{M}}

\newcommand{\msun}{M_{\odot}}
\woctitle{Tidal Disruption and AGN outburst}
\begin{document}
\title{Challenges in the modelling of tidal disruption events lightcurves}
%
%

\author{Giuseppe Lodato\inst{1}\fnsep\thanks{\email{giuseppe.lodato@unimi.it}}
}

\institute{Dipartimento di Fisica, Universit\'a degli studi di Milano
          }

\abstract{%
In this contribution, I review the recent developments on the modelling of the lightcurve of tidal disruption events. Our understanding has evolved significantly from the earlier seminal results that imply a simple power-law decay of the bolometric light curve as $t^{-5/3}$. We now know that the details of the rise to the peak of the lightcurve is determined mainly by the internal structure of the disrupted star. We also have improved models for the disc thermal emission, showing that in this case the decline of the luminosity with time should be much flatter than the standard $t^{-5/3}$ law, especially in optical and UV wavelengths, while the X-ray lightcurve is generally best suited to track the bolometric one. Finally, we are just starting to explore the interesting general relativistic effects that might arise for such events, for which the tidal radius lies very close to the black hole event horizon. In particular, I describe here some possible evidences for relativistic Lense-Thirring precession from the light curve of the event Swift J1644. 
}
\maketitle 
\section{Introduction}
\label{intro}

The theory of tidal disruption events (TDE) by supermassive black holes has been developed since the early '80s with the seminal work of Carter and Luminet \cite{carter82,carter83} Lacy, Townes and Hollenbach \cite{lacy82}, Rees \cite{rees88} and Phinney \cite{phinney89}. However, for several years little progress has been made, except for some important contribution from the numerical point of view \cite{evans89,ayal2000} and on the analysis of the accretion regime \cite{cannizzo90}. 

In recent years, flares from the nuclei of quiescent galaxies, detected either in X-rays \cite{komossa99,esquej08,cappelluti09} and in the UV \cite{gezari08,gezari09,gezari12} have been interpreted as arising from TDEs, essentially by a simple comparison of the light curve of the event to the expected $t^{-5/3}$ law \cite{rees88}. This has prompted several new theoretical investigations that aimed at clarifying the conditions under which one should really expect such a power-law decline in the luminosity of the event. 

The field has further expanded once it has been realised observationally that such events might produce powerful relativistic jets \cite{bloom11,cenko12}, whose modelling clearly needs to go beyond the simple estimate of the dynamics of the infalling debris and the disc thermal emission \citep{giannios11}. Additionally, the larger scale outflows that might be connected with the early super-Eddington fallback phase is expected to produce a characteristic spectroscopic signature \cite{strubbe09,strubbe11}. 

In this framework, modelling the dynamics of the debris is still a key issue in order to understand the time evolution of the accretion rate onto the black hole, which is an essential input parameter for any detailed model of the system emission properties. In this contribution I will first describe the basic results and the standard theory behind it and I will then focus on a number of recent developments that have somewhat changed the basic expectations of the simplest models. 

\section{Light curve modelling}
\label{sec-1}

\begin{figure*}
\centering
\includegraphics[width=0.4\textwidth,clip]{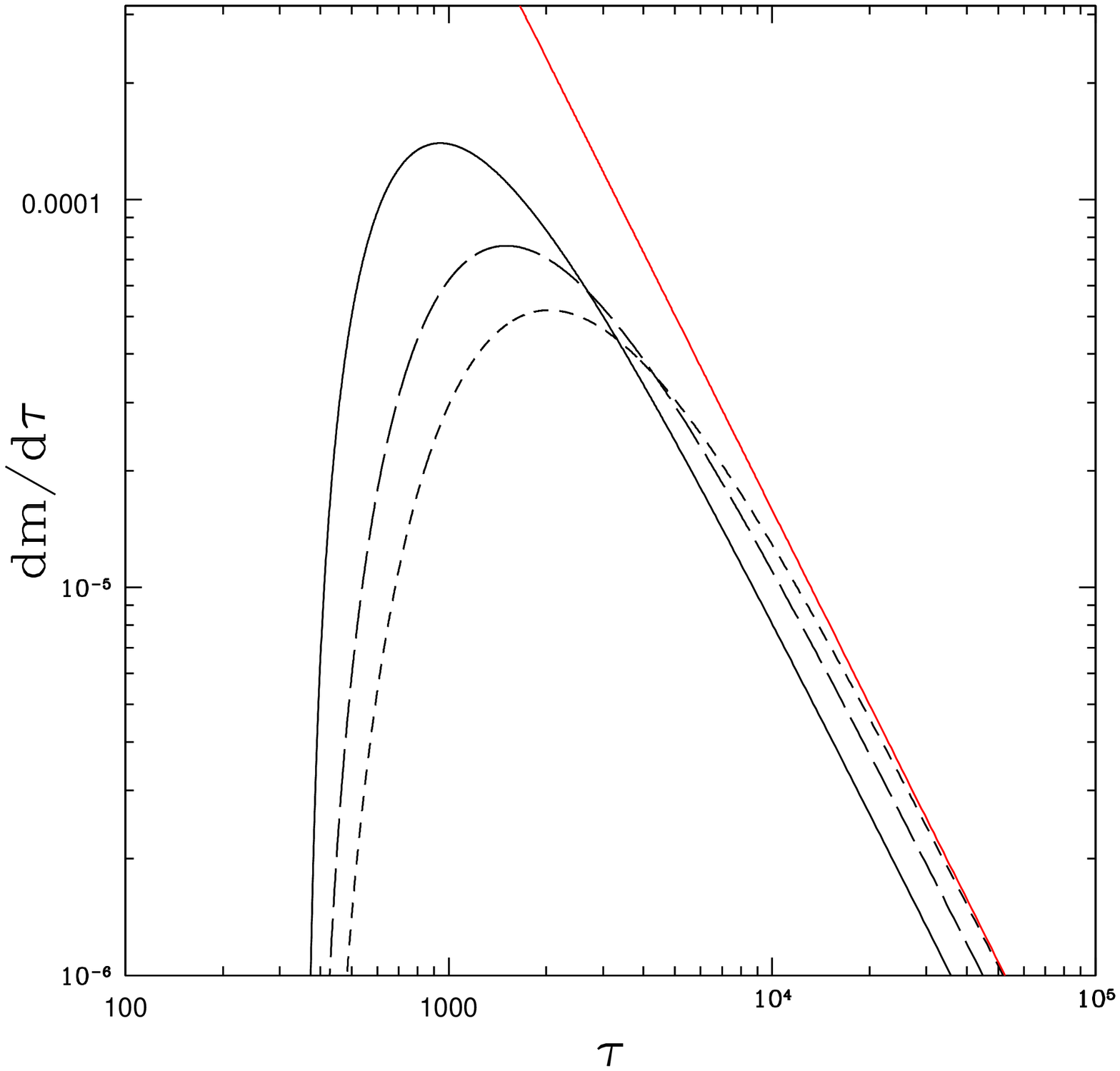}
\includegraphics[width=0.4\textwidth,clip]{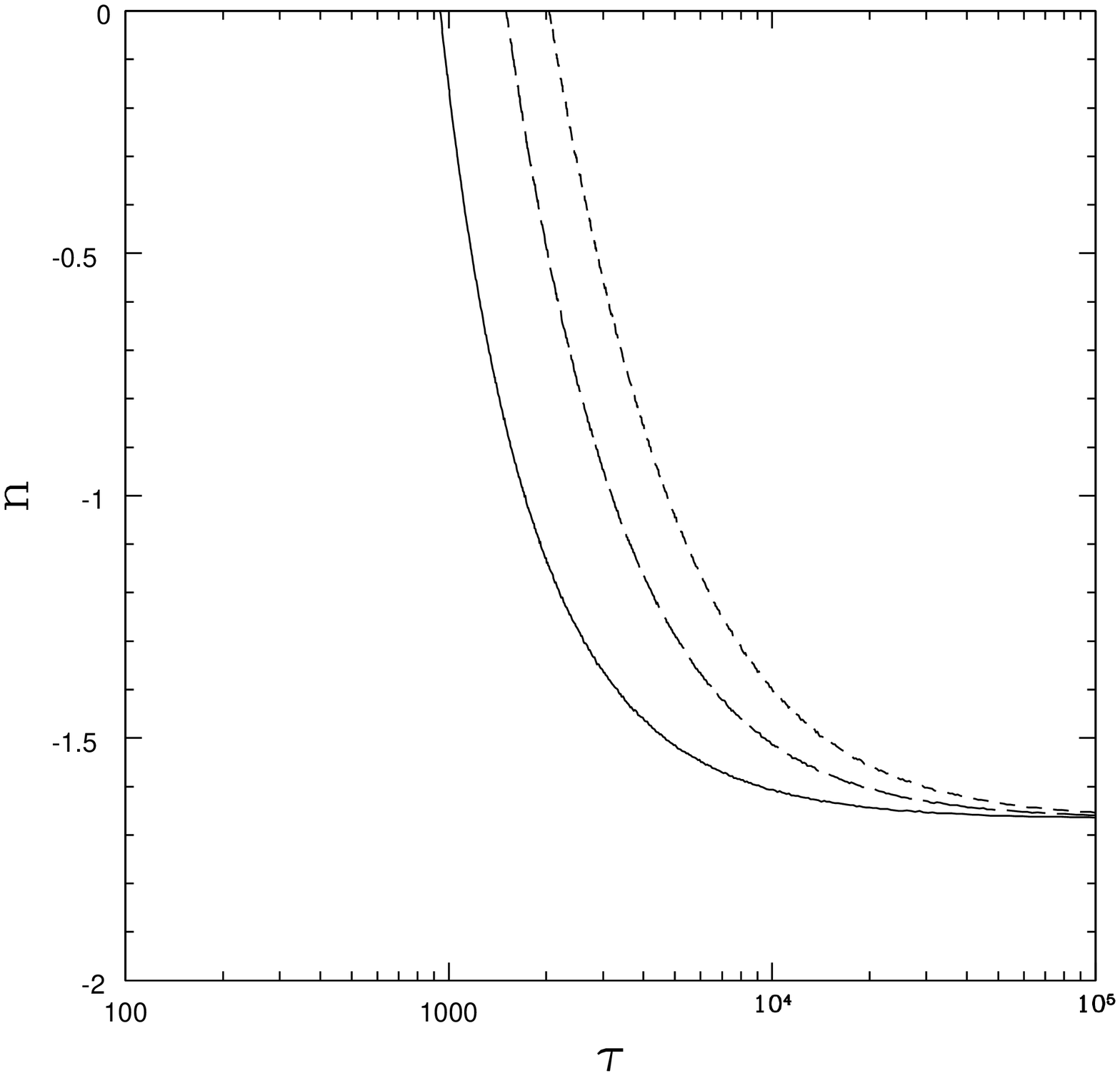}
\caption{Left: fallback rate as a function of time for the disruption of three polytropic stars, with progressively decreasing values of $\gamma=5/3$ (solid), $\gamma=1.4$ (long-dashed), and $\gamma=4/3$ (short-dashed). Right: corresponding instantaneous value of the logarithmic derivative of $\mdot(t)$. More compressible, and centrally concentrated stars, have a more gentle rise to the peak, and the $t^{-5/3}$ regime is reached later. From \cite{LKP09}.}
\label{fig:LKP1}       
\end{figure*}

\subsection{Classic results}

The tidal radius beyond which stellar self-gravity is not able to counteract the black hole's tidal field and keep the star together is defined as:
\begin{equation}
R_{\rm t}=\left(\frac{M_{\rm h}}{M_*}\right)^{1/3}R_*,
\end{equation}
where $M_{\rm h}$ is the black hole mass, $M_*$ the stellar mass and $R_*$ the stellar radius. Note that for typical parameters this is not much larger than the black hole's Schwarzschild radius $R_{\rm S}$:
\be
\frac{\rt}{\rs}\approx 23M_6^{-2/3},
\ee
where $M_6=\mh/10^6M_{\odot}$, which implies that black holes with masses above $\approx 10^8M_{\odot}$ are not expected to produce any TDE. It should be noted however that the above limit is only valid in Newtonian dynamics. General relativistic calculations have shown \cite{kesden12} that highly spinning black holes are able to produce stellar disruption of solar type stars for masses well above $10^8\msun$. This is a first example that shows how important general relativistic effects can be in describing TDEs, and thus be a powerful probe of black holes properties otherwise difficult to measure, such as their spin. 

Often, the orbit of the incoming star is described as a simple parabolic orbit, with a pericenter distance $\rp$ close to the tidal radius. One can define a ''penetration factor'' $\beta=\rt/\rp$, such that when $\beta<1$ tidal disruption should not occur. Since the details of the light curve essentially depend on the conditions at the tidal radius (see below), in the following we will only restrict to the case where $\beta=1$ so that the tidal radius coincides with the pericenter of the stellar orbit. 

After disruption, the stellar debris are launched into very eccentric orbits (with $1-4R_*/\rp<e<1$) with a spread of orbital energies, and gradually return to pericenter, where they circularise and form an accretion disc at $R_{\rm circ}\approx 2\rp$. During this circularisation phase the debris lose a fair fraction of their energy, which might be in principle detectable. Note, however, that the ratio between the luminosity released during circularisation and that released by the subsequent accretion of the debris is expected to be
\be
\frac{L_{\rm circ}}{L_{\rm acc}}\approx\frac{R_{\rm ISCO}}{R_{\rm circ}}\approx 0.06\beta M_6^{2/3},
\ee
and is thus generally neglected in most calculations. For large black hole masses, or for deeply penetrating events, this contribution might become significant. 

The fallback rate of the debris is easily derived from the the distribution of their orbital energies $\mbox{d}M/\mbox{d}E$ and Kepler's third law (the effects of general relativity can also been included, \cite{kesden12b}):
\be
\mdot_{\rm fb}=\frac{(2\pi G\mh)^{2/3}}{3}\frac{\mbox{d}M}{\mbox{d}E}t^{-5/3},
\ee
which, in the limit that the energy distribution of the debris is flat, gives the standard $t^{-5/3}$ decay law, almost universally associated with this type of events \cite{rees88}. A simple order of magnitude estimate for $\mbox{d}M/\mbox{d}E$ is obtained by assuming that the debris are distributed uniformly within the potential energy spread across a stellar radius:
\be
\frac{\mbox{d}M}{\mbox{d}E} \approx\frac{\rt^2}{2G\mh}\frac{M_*}{R_*}.
\ee

\subsection{Recent developments}

\subsubsection{Effects of stellar structure}

The theory described above highlights the fundamental role played by the distribution of orbital energies of the debris. Clearly the assumption that it is flat is a first crude approximation, and the actual shape of the distribution, and hence the actual time dependence of the fallback rate needs to be computed. The issue has been initially treated from a purely numerical point of view by Evans and Kochanek \cite{evans89}, who, somewhat surprisingly, appeared to confirm that indeed the hypothesis of a flat distribution originally made for simplicity by Rees \cite{rees88} was correct. More recently, Lodato, King and Pringle \cite{LKP09} have revisited the problem both from an analytical and a numerical point of view. 

\begin{figure}
\centering
\includegraphics[width=0.4\textwidth,clip]{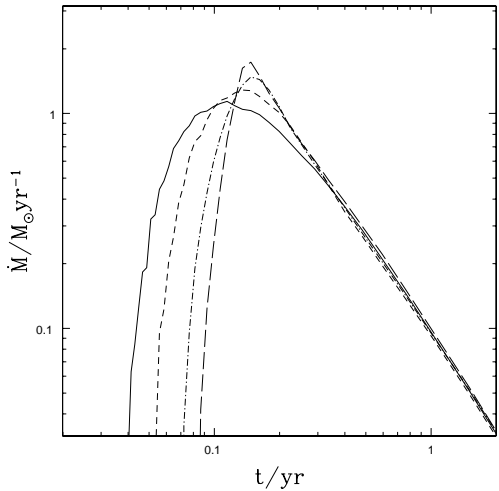}
\caption{Expected fallback rates for the disruption of polytropic stars, with progressively increasing values of $\gamma=1.4$ (solid), $\gamma=1.5$ (short-dashed), $\gamma=5/3$ (dot-dashed) and $\gamma=1.8$ (long-dashed), scaled to solar values. More centrally concentrated stars (with a lower $\gamma$) have a more gentle rise to the peak and peak at an earlier time with respect to more incompressible models. From \cite{LKP09}.}
\label{fig:LKP2}       
\end{figure}

The analytical model is based on the assumption that the structure of the disrupted star is essentially unperturbed until it reaches the tidal radius. The assumption is clearly approximate, but is needed in order to compute analytically the resulting energy distribution, which then needs to be compared to numerical simulations. The big advantage of such analytical model is that it  essentially allows to compute the debris energy distribution, and hence the fallback rate, starting from the internal density profile of the unperturbed disrupted star:
\be
\frac{\mbox{d}M}{\mbox{d}E}= \frac{2\pi\rt^2}{G\mh}\int_r^{R_*}\rho(r')r'\mbox{d}r',
\ee
where $r=E\rt^2/G\mh$. The main result of this theory is that more compressible stars, which are more centrally concentrated, have a more gentle rise to the peak, in the sense that the logarithmic gradient of the light curve reaches the expected $-5/3$ value at later times (see Fig. \ref{fig:LKP1}). 

Lodato, King \& Pringle \cite{LKP09} have also tested this model numerically, by running SPH simulations of the disruption of simple polytropic stars, with $\beta=1$. These numerical tests reproduce to a large extent the predictions of the model, hence confirming that the analytical model is a good first approximation. However, there are two important changes to be taken into account: (a) the effective reduction of the stellar self-gravity as it approaches the tidal radius determines an inflation of the star itself and (b) shocks, that develop preferentially in more compressible stars, tend to push matter at the edges of the energy distribution, thus further modifying the fallback rate. The net result of these two effects is to delay the time of the peak for the less centrally concentrated stars (see Fig. \ref{fig:LKP2}). Such results have also been confirmed by later investigations of the same problem \cite{guillo}.

Form an observational point of view, the models above have been used to describe the UV emission of several TDE candidates \cite{gezari09}. In particular, very recently Gezari et al \cite{gezari12} have used the arguments above to conclude that a peculiar TDE flare can be attributed to the disruption of the helium core of a red giant, whose envelope had been previously removed (possibly by the same tidal mechanism that later caused its complete disruption). All such analyses assume that the UV an optical light curve is proportional to the fallback rate. Whether such assumption is correct should be discussed in greater detail (see next section). 

It is worth noting that the above investigations have only considered the case where the pericenter of the orbit coincides with the tidal radius, and hence $\beta=1$. How large is the energy spread of the debris for more penetrating encounters? Is it the relatively small spread as determined at the tidal radius or does it rather reflect the conditions at pericenter? The problem has been first investigated by Sari et al \cite{sari10} in the related context of tidal disruption of binary systems by a supermassive black hole, and later confirmed also in the context of stellar disruption \cite{guillo}. The conclusion is that the energy spread is determined by the conditions at the tidal radius, and is thus much smaller than what would be naively predicted if one computes it based on the conditions at the pericenter. 

\begin{figure}
\centering
\includegraphics[width=0.4\textwidth,clip]{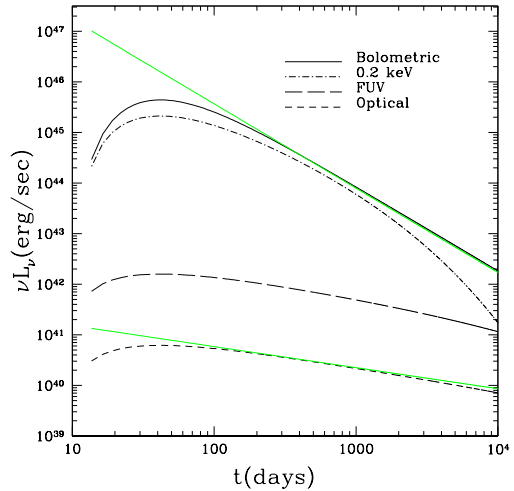}
\caption{Bolometric, X-ray (0.2 keV), FUV and optical light curves of the tidal disruption of a solar mass star by a $10^6M_{\odot}$ black hole. Only the disc thermal emission is shown. The green lines indicate a $t^{-5/3}$ and a $t^{-5/12}$ decline, respectively. From \cite{LR11}.}
\label{fig:LR}       
\end{figure}

\subsubsection{Modelling the thermal disc emission}

Up to now, I have just considered the modelling of the fallback rate, which shows, at least after the peak of the emission, the typical $t^{-5/3}$ decline with time. How does this convert into luminosity? 

First of all, note that the fallback rate is the rate at which the debris return to pericenter and circularise, forming an accretion disc. As noted above, most of the luminosity associated with a tidal disruption event is released as the debris accrete from the circularisation radius onto the black hole through an accretion disc. It is thus the accretion rate $\mdot$ onto the black hole that determines the luminosity of the flare. This in general is not equal to the fallback rate $\mdot_{\rm fb}$, unless the viscous timescale in the disc is much smaller than the timescale related to the fallback of the debris. We will discuss this condition more in detail in the following section. For the moment, let us just make the usual and simplifying assumption that $\mdot=\mdot_{\rm fb}$. 

In this case, the fallback rate essentially sets the scale for the bolometric luminosity of the system, trhough $L=\eta\mdot_{\rm fb} c^2\propto t^{-5/3}$, where $\eta$ is the accretion efficiency. However, the light curves at specific wavelengths might strongly differ for the above behaviour, given that the system also cools down significantly while fading. The issue has been studied in detail by Lodato \& Rossi \cite{LR11}, who describe the time evolution of the thermal disc emission at various wavelengths. It can be thus demonstrated that the light curve, at any given wavelength, should go through three different phases. (1) Initially, when the disc is very hot, the observed wavelength might fall on the Rayleigh-Jeans tail of the blackbody spectrum produced by the disc. The luminosity $L_{\nu}$ in this case is thus proportional to the temperature $L_{\nu}\propto T_{\rm disc}\propto \mdot_{\rm fb}^{1/4}\propto t^{-5/12}$ (note that this specific result has also been previously suggested by \cite{strubbe09}). (2) As the disc cools, the observed wavelength might correspond roughly to the peak of the blackbody spectrum. It is in this phase that one should expect the single wavelength light curve to decline as $t^{-5/3}$, as the bolometric light curve. (3) Finally, once the observed wavelength moves to the Wien part of the spectrum, the light curve will start to rapidly fade off exponentially. The first of these three phases might be absent if the observed wavelength is high, for example in X-rays. These three behaviours can be clearly seen in Fig. \ref{fig:LR}. 

The above results thus indicate that X-ray observations are most likely to display the canonical $t^{-5/3}$ decline, while the UV and optical light curve should display a much shallower light curve. Such a shallower decay in optical and UV has been observed in the case of Swift J2058 \cite{cenko12}. 

\begin{figure}
\centering
\includegraphics[width=0.5\textwidth,clip]{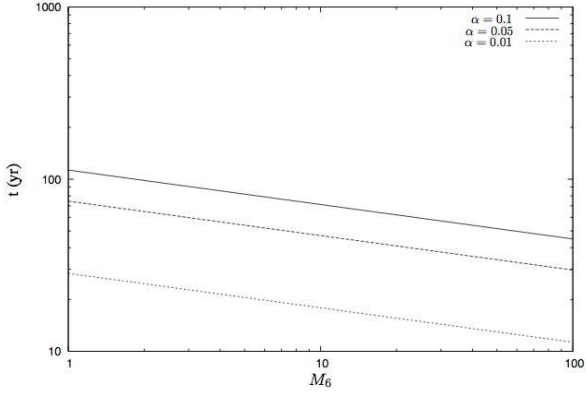}
\caption{The expected timescale for the transition from a fallback dominated accretion regime into a viscosity dominated regime, as a function of the black hole mass for various choices of $\alpha$.}
\label{fig:trans}       
\end{figure}

\subsubsection{Fallback versus viscosity dominated accretion}

As mentioned above, one key assumption of the traditional modelling of TDE is that the viscous timescale in the disc is much smaller than the return timescale of the debris. In this way, the disc can be described by a series of steady state configurations with varying $\mdot$, at each time equal to the incoming fallback rate $\mdot_{\rm fb}$. At the opposite end, some models \cite{cannizzo90} have considered the behaviour of a system where the fallback material forms a narrow ring, which at later times starts accreting onto the black hole. Such models are appropriate only when the viscous timescale in the disc is always much longer than the fallback timescale. 

In reality, neither model is correct. At earlier times, when the accretion rate is close to Eddington and the disc is hot and thick, it is expected that the viscous timescale is indeed much shorter than the fallback time, while at later times, once the accretion rate has fallen off significantly and the disc has cooled down, the ordering of the relevant timescales should be reversed, and the TDE should transition from a \emph{fallback dominated accretion} regime, where the accretion rate coincides with $\mdot_{\rm fb}$, into a \emph{viscosity dominated accretion regime}, where it is the internal disc viscosity that determines the time evolution of $\mdot$. 

\begin{figure}
\centering
\includegraphics[width=0.5\textwidth,clip]{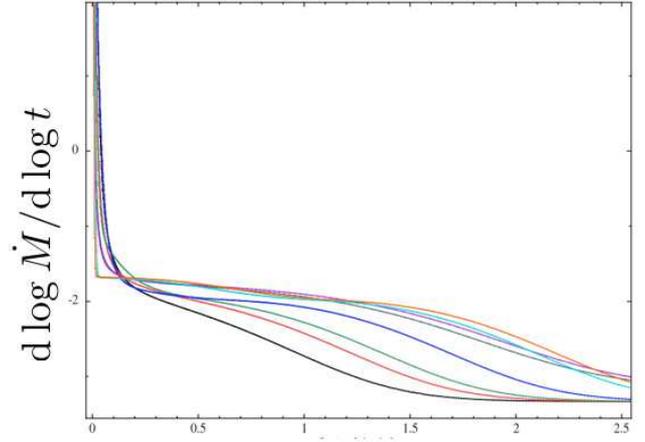}
\caption{Evolution of the instantaneous power law index of the light curve as a function of time for several choices of $t_{\rm trans}$ (decreasing from top to bottom). The $x$-axis shows $\log(t/t_{\rm min})$, where $t_{\rm min}$ is the return time of the most tightly bound debris, marking the beginning of the flare. For the case indicated with a black line (explored here for completeness), $t_{\rm trans}$ is comparable to $t_{\rm min}$ so that the $t^{--5/3}$ plateau does not occur. }
\label{fig:trans2}       
\end{figure}

The timescale for such transition can be easily computed based on the following argument. The timescale over which the fallback rate evolves is:
\be
t_{\rm fb} = \frac{M_*}{\mdot_{\rm fb}} \propto t^{5/3}.
\ee
On the other hand, the viscous timescale in the disc is proportional to $(H/R)^{-2}$ (where $H$ is the disc thickness). At the high accretion rates predicted for TDE the disc is likely to be radiation pressure dominated and thus $H/R\propto \mdot/\mdot_{\rm Edd}$. We thus have:
\be
t_{\nu} = \frac{1}{\alpha}\left(\frac{H}{R}\right)^{-2}\propto \mdot_{\rm fb}^{-2} \propto t^{10/3},
\ee
where $\alpha$ is the disc viscosity coefficient and thus grows with time much faster than $t_{\rm fb}$. Thus, even if at small times $t_{\nu}\ll t_{\rm fb}$, at later times we should expect a transition. The time at which the two timescales become equal is:
\be
t_{\rm trans} = 110\left(\frac{\alpha}{0.1}\right)^{3/5}\left(\frac{\mh}{10^6M_{\odot}}\right)^{-1/5}
\left(\frac{M_*}{M_{\odot}}\right)^{3/2}\left(\frac{R_*}{R_{\odot}}\right)^{-3/2}\mbox{yr},
\ee
which shows that for most cases this transition actually occurs after several decades from the disruption, once the accretion rate has dropped too much to be significant (cf. \cite{strubbe09}). Figure \ref{fig:trans} shows this timescale as a function of the black hole mass for several values of $\alpha$. 

Detailed time-dependent calculations of this effect (Lodato, in preparation) show that significant departures (to a 20\% level) to the canonical $t^{-5/3}$ decline can occur already at a fraction of $t_{\rm trans}$ as small as 0.1. Figure \ref{fig:trans2} shows the results of one such calculations for various values of $t_{\rm trans}$. In particular we plot the instantaneous power-law index of the light curve as a function of time. One can see that after a plateau at $-5/3$, the light curve steepens significantly, once accretion through the disc slows down.

\section{Modelling Swift J1644: lightcurve and precession}

Swift J1644 has been one of the most spectacular TDE reported to date, showing for the first time that powerful jets can be associated with such events \cite{bloom11,levan11,burrows11,zauderer11}. A detailed modelling of the spectral energy distribution can reveal what are the main emission mechanisms for the jet emission \cite{bloom11}. An analysis of the time evolution of the light curve can, on the other hand, give us clues on the dynamical features of the event. Such a dynamical analysis has been attempted \cite{CTL11} based on the early light curve. 

\begin{figure}
\centering
\includegraphics[width=0.5\textwidth,clip]{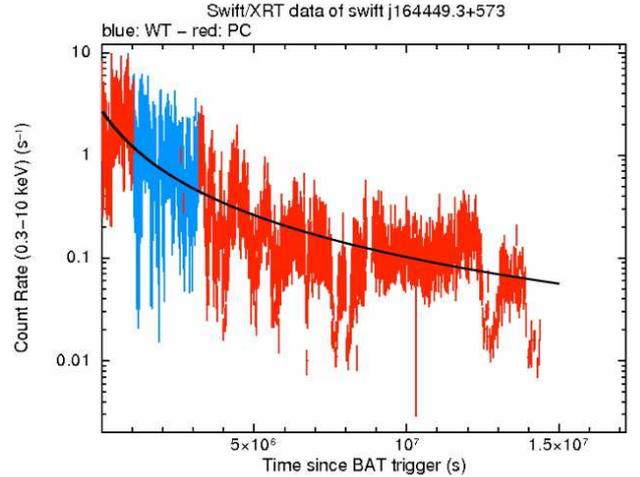}
\caption{The blue and red curves show the XRT light curve of Swift J1644, while the black line is a rough fit to it, using a simple $t^{-5/3}$ law.}
\label{fig:swift}       
\end{figure}

We can attempt here a new analysis based on the light curve during the first half a year of activity. Figure \ref{fig:swift} shows the XRT count rate (red and blue lines) for the event during the first six months. The system is very variable, but a simple $t^{-5/3}$ can roughly reproduce the data (black line). This has been obtained by assuming that the jet luminosity scales with the fallback rate:
\be
L_{X}(t) = f\mdot_{\rm fb}c^2,
\ee
where $f$ is a scaling factor that might not be too different than unity, since while it is true that only a fraction of the accretion energy is expected to be released from the jet, one should also consider that the X-ray luminosity is boosted by relativistic beaming. We further assume the following behaviour of $\mdot_{\rm fb}$ with time:
\be
\mdot_{\rm fb}(t) = \mdot_{\rm peak}\left(1+\frac{t}{t_{\rm min}}\right)^{-5/3},
\ee
where $t_{\rm min}$ is the return time of the most bound debris and one should remember that in this case $t=0$ marks the beginning of the outburst (the BAT trigger) rather than the pericenter passage of the star, as before. 

\begin{figure}
\centering
\includegraphics[width=0.4\textwidth,clip]{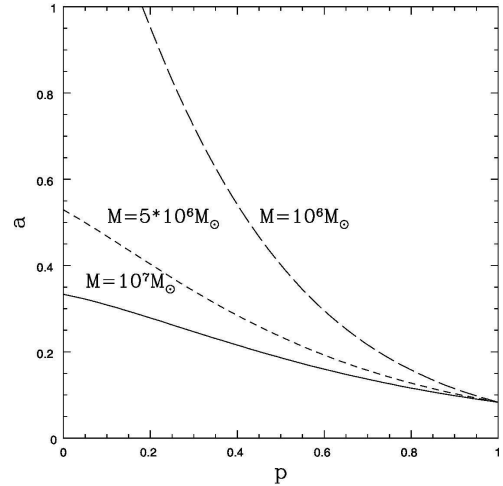}
\caption{The spin parameter $a$ of the black hole in Swift J1644, as implied by assuming that the 2.7 days variability of the X-ray lightcurve is due to disc precession caused by the Lense-Thirring effect. Here $p$ is the power-law index of the surface density profile (small $p$ implying flat profiles). The values of the assumed black hole mass are indicated.}
\label{fig:spin}       
\end{figure}

This comparison allows us to estimate the following parameters: $t_{\rm fb}\approx 19$ days, $f\mdot_{\rm peak}c^2\approx 4~10^{46}$ ergs/sec, $\mh\approx 2f~10^6M_{\odot}$, $M_*\approx 0.1M_{\odot}/f$. We can thus see that the parameters estimated in this way are in line with the expectations from such event, for an $f$ factor not too different from unity. 

Swift J1644 might also show some additional interesting features connected with relativistic effects. The X-ray lightcurve indeed shows an interesting 2.7 days periodicity \cite{burrows11} that might be connected with jet precession. Indeed, Lei et al \cite{lei12} have suggested that such modulation is linked to precession induced by the Bardeen-Petterson effect due to a warping of the disc caused by the torque associated with the spin of the accreting black hole. However, this is probably not the case because: (a) the Bardeen-Petterson effect does not lead to a steady precession. The inner disc rapidly gets aligned with the black hole spin, and the final realignment of the latter with the outer disc occurs on a timescale comparable with the precession timescale \cite{LP06}; (b) the Bardeen-Petterson effect occurs for very thin and viscous discs, which is not the case for the almost-Eddington accretion rates involved in TDE. 

In reality, though, a steady precession of the disc is expected in the case of thick configurations (see also \cite{stone12}), where the warp propagates in the disc as a wave, provided that the disc radial extent is short enough (which is also expected in this case). The precession frequency is obtained simply by balancing the overall torque produced in the disc by the frame-dragging effect, and the disc inertia:
\be
\Omega_{\rm p} = \frac{\int\Omega_{\rm LT}(R)L(R)R\mbox{d}R}{\int L(R)R\mbox{d}R},
\ee
where $\Omega_{\rm LT}$ is the Lense-Thirring precession rate \cite{LP06}. Since $\Omega_{\rm LT}$ depends on the black hole spin parameter $a$, one can use the known precession rate $\Omega_{\rm p}$ to infer the black hole spin. In particular, we can assume that the disc surfece density has a simple power-law dependence on radius $\Sigma(R)\propto R^{-p}$, where $p$ is a free parameter. Figure \ref{fig:spin} shows the implied value of $a$ as a function of $p$ for several values of the black hole mass. We thus see that, unless the black hole mass is small and the surface density profile is very flat, we estimate a rather low value for $a\sim 0.2-0.4$ with a relatively weak dependence on the free parameters.

\section{Conclusions}

The modelling of the light curve of tidal disruption events has moved significantly from the earlier standard results implying a $t^{-5/3}$ decline. We now know much better how to model the initial rise of the lightcurve, and how it is connected with the internal structure of the disrupted star. We begin to model in greater details the specific lightcurves at various wavelengths, although further work is required in this context to understand why in some cases the observed UV lightcurves appear not to cool down with time as expected from a thermal disc emission. Finally we are just now starting to understand how relativistic effects (for example connected with the black hole spin) can affect the light curve of the system. In this respect, TDE offer a unique way to probe relativistic effects, given that the tidal radius is often very close to the black hole event horizon. Further specific modelling of such effects is going to be essential in order to correctly interpret upcoming observations. 

%
\bibliography{lodato}

\end{document}